\documentclass[10pt,conference]{IEEEtran}

\IEEEoverridecommandlockouts
\usepackage{graphicx}
\usepackage{amssymb}
\usepackage{algorithm}
\usepackage{stackrel}
\usepackage{algpseudocode}
\usepackage{textcomp}
\usepackage{listings}
\usepackage[usenames,dvipsnames]{xcolor}
\usepackage[skip=2pt]{subcaption}
\usepackage[skip=2pt]{caption}
\usepackage{amsmath}

\usepackage{paralist}
\usepackage[bookmarks=false]{hyperref}
\usepackage{lipsum}
\usepackage[normalem]{ulem}
\usepackage{array}
\usepackage{multicol}
\usepackage{enumitem}
\usepackage{etoolbox}
\usepackage{varwidth}
\usepackage{xspace}

\usepackage{multirow}

\makeatletter
\preto{\@verbatim}{\topsep=5pt \parsep=0pt }
\makeatother
\newcolumntype{L}[1]{>{\raggedright\let\newline\\\arraybackslash\hspace{0pt}}m{#1}}
\newcolumntype{C}[1]{>{\centering\let\newline\\\arraybackslash\hspace{0pt}}m{#1}}
\newcolumntype{R}[1]{>{\raggedleft\let\newline\\\arraybackslash\hspace{0pt}}m{#1}}

\newcommand{\tinysection}[1]{\noindent \textbf{#1.}~}

\usepackage{graphicx}
\usepackage{times}
\usepackage{fancyhdr}
\usepackage{balance}

\definecolor{gray}{rgb}{0.4,0.4,0.4}

\ifCLASSINFOpdf
\else
\fi
\begin{document}

\title{Detecting Data Leakage from Databases \\on Android Apps with Concept Drift}

\author{\IEEEauthorblockN{G\"{o}khan Kul, Shambhu Upadhyaya, Varun Chandola}
\IEEEauthorblockA{
%Department of Computer Science and Engineering\\
University at Buffalo, SUNY\\
Buffalo, New York 14260--2500\\
\{gokhanku, shambhu, chandola\}@buffalo.edu}
}


%Telephone: (800) 555--1212, Fax: (888) 555--1212}


\maketitle

\begin{abstract}
% !TEX root = ../paper.tex
Mobile databases are the statutory backbones of many applications on smartphones, and they store a lot of sensitive information. However, vulnerabilities in the operating system or the app logic can lead to sensitive data leakage by giving the adversaries unauthorized access to the app’s database. In this paper, we study such vulnerabilities to define a threat model, and we propose an OS-version independent protection mechanism that app developers can utilize to detect such attacks. To do so, we model the user behavior with the database query workload created by the original apps. Here, we model the drift in behavior by comparing probability distributions of the query workload features over time. We then use this model to determine if the app behavior drift is anomalous. We evaluate our framework on real-world workloads of three different popular Android apps, and we show that our system was able to detect more than 90\% of such attacks.
\end{abstract}


% \fi
\IEEEpeerreviewmaketitle

\section{Introduction}
% !TEX root = ../paper.tex
%\subsection{Motivation}

Google Android OS usage has grown substantially over the past years and reached 85\% of the market share in smartphones by the first quarter of 2017~\cite{marketshare}.
Unlike its competitors, Android OS is open source, and used by many hardware vendors on their smartphones.
However, this flexibility comes with the cost of hardware interface (a.k.a. firmware) development by the vendors.
This results in some smartphone models getting out-of-date, and not supported by the new versions of the Android OS, hence, not being able to get the latest security updates.
Based on this limitation, application (app) developers may need to employ their own defense mechanisms if their apps get affected by vulnerabilities.
%Instead of depending on the OS, we can relay this problem from the app developers.
Apps can release new versions for each Android version even though the OS support is terminated.

On smartphones, both the operating system and the apps contain a lot of sensitive information that is subject to various threats~\cite{androidsecurity2016, hassanshahi2017android, jain2017sniffdroid, bhandari2017sneakleak}.
Some of these threats exploit vulnerabilities in the OS, or take advantage of the flexible app development capabilities of Android which expose the access credentials of databases dedicated to the apps.
Our solution targets apps that store sensitive information in their databases, and it is based on monitoring database access activity 
%within the organization
of the users~\cite{brodsky2015dam}.
This approach provides a flexible solution that can be employed by the apps themselves, the database system, or the operating system. It enables monitoring unusual activity, and take action against potential threats.
The advantage is to be able to observe all, including permissible, activity.
Since a smartphone is essentially designed to be a personal device, the underlying database is designed to be owned and used only by the app.
%Hence, the database access policies cannot be used to protect data.
Android database security features are designed to prevent all other users except the app itself to use the database.

We can employ the monitoring mechanism to depend on detecting anomalies in the app's database usage. The basic pathway for a common monitoring mechanism is to:
(1) extract relevant features that reflect the user behavior,
(2) cluster similar actions,
and (3) find outlier actions, which appear to be a very effective approach~\cite{Kamra2007SyntaxBased}.
However, attacks exploiting vulnerabilities can use the app's query generator to create queries, which would result in issuing the same query templates to the database, therefore, these queries cannot be considered as outliers with this method.
Furthermore, smartphone apps can change over time with small modifications such as updates, and force the users to accomplish the same task in a different way~\cite{maggi2009protecting}.
Also, the users can get more proficient with the use of app in time, or their interests can shift over time.
Both of these cases would require the deployed security mechanisms to adapt to the change.

%https://www.ukp.tu-darmstadt.de/fileadmin/user_upload/Group_TRUST/PubsPDF/xmandroid.pdf

\begin{scenario}
Let us consider a photographer (let's call him Jason) using Instagram profile as a portfolio.
During the first month of the account opening, he aggressively posts his existing photographs, and builds a portfolio.
He then uses the Instagram profile to answer the questions, and communicate with the potential customers.
A few weeks later, Instagram introduces a new feature called ``Story'', which enables users to post short videos and pictures that disappear automatically after 24 hours.
Considering taking quality photographs that attract customers takes time and energy, he less frequently posts photos, and he starts to post stories while on the job, to keep the interest of his followers.
\end{scenario}

Jason's activity results in Instagram web services to query the database in three different areas of the database:
(1) Permanent photo storage, (2) Messaging storage, and (3) Story storage.
However, in the first month, his activity constitutes mostly inserting to the permanent photo storage, and increasing read and write access to the messaging storage. After that time period, the queries generated for his activity target mostly inserting to and deleting from the story storage while maintaining a similar workload on messaging, and less frequently generating insert queries to the permanent photo storage.
Although both of these behavior shifts are expected, the shifts from normal profile activity can be identified as an outlier if the monitoring system is not designed to adapt to the changes, resulting in increased false positives.

A work-around for this problem could be re-training with more recent data when the system starts to return false positives.
However, an adaptive system can tolerate this change, therefore, reduce false positives, and eliminate  the need for retraining.
We observe that various experiments reported in the literature do not consider the changes that happen over time, and the synthetic datasets used for the experiments do not reflect any behavior changes.

In this paper, we bridge this gap by constructing a user behavior model which explicitly models \textit{temporal behavior drift}.
We focus on detecting data breaches against an app's database to scope down the problem.
%We argue that this approach is effective in reducing the false positive rates while achieving the same or better performance with regards to true positive rates, compared to the ordinary threshold based approach.
In our experiments, we utilize real-world query logs of three different apps to understand the behavior drift, and validate the effectiveness of our proposed system. 

%There are two ways to detect these threats: (1) misuse detection, and (2) anomaly detection.

%Furthermore, we will analyze the possible actions of the users to model the threat they pose for the system. \todo{Talk about the complexity analysis and threat modeling approach.} Hence, we will be able to design the defense mechanism in order to counter the moves of the insider with minimal human intervention. We will use the CERT insider threat dataset~\cite{glasser2012dataset} to analyze normal human behavior and the nature of insider attacks and then investigate red-teaming strategies for a SQL query log dataset collected from a national bank's database systems~\cite{kul2016ettu}.

%\subsection{Our Techniques}

To identify the behavior drift, and determine if there is a possible data leakage in a user's workload, we define four basic steps. We first observe and process every SQL query issued to the DBMS. We extract the relevant features of the query considering which part of the database the user is accessing with that particular query. Second, we construct user profiles by accumulating the extracted features for each user for a given period of time. The distribution of the features harvested from these queries constitutes a user profile. Third, we identify the constant drift in behavior by observing the changes in the distribution of features the users utilize. Lastly, we analyze the behavior change, and determine outliers by detecting drastic changes in the drift. Namely, we consider the temporal behavior drift of each user to set an adaptive threshold.

In summary, the contributions of this paper are:
(1) Building a threat model that leads to sensitive data leakage from smartphones,
(2) Introducing \emph{behavioral drift} in user activity,
and (3) Providing a defense strategy against the modeled threat.
Additionally, to the best of our knowledge, our work is the first study that uses a large real--world SQL query trace to validate the results.

This paper is organized as follows.
%We start by summarizing the fundamental database concepts that we will use in this work in Section~\ref{sec:preliminaries}, and 
we start by introducing our system Query Workload Auditor (QWA) and the threat model it addresses in Section~\ref{sec:threatmodel}.
%we present the related work in Section~\ref{sec:relatedwork}.
We then explain the techniques that we use to create user profiles in Section~\ref{sec:usermodeling}.
Section~\ref{sec:experiments} presents the experiments performed to show the effectiveness of the system, and we discuss the shortcomings and potential improvements over the proposed system in Section~\ref{sec:discussion}.
We present the related work in Section~\ref{sec:relatedwork} and discuss our work in Section~\ref{sec:discussion}.
Finally, we conclude and discuss our plans to improve our system in Section~\ref{sec:conclusions}.

%\section{Related Work}
%\label{sec:relatedwork}
%\input{sections/3-relatedwork.tex}

%\section{Preliminaries}
%\label{sec:preliminaries}
%\input{sections/2-preliminaries.tex}

\section{QWA}
\label{sec:threatmodel}
% !TEX root = ../paper.tex

Query Workload Auditor (QWA) aims to provide a fix
%on the application level
to the vulnerabilities that can lead to data leakage from the app's database, using the data access patterns of each specific app. In this section, we first discuss recent vulnerabilities that can cause the described threat in Section~\ref{sec:vulnerabilities}, the threat model we address in Section~\ref{sec:threat}, and the architecture of the described system in Section~\ref{sec:architecture}.

\subsection{Vulnerabilities}
\label{sec:vulnerabilities}

There are a number of vulnerabilities in the Android OS reported in the literature, most of which are fixed~\cite{zhang2014appsealer, yan2012droidscope}. In this paper, we focus on recent vulnerabilities reported which can allow attackers to steal sensitive information from the app's database.

\tinysection{Janus Vulnerability} Android Application Package (APK) file format is used to distribute Android apps. It is essentially a compressed ZIP file that is structured to be recognized as an app distribution package by Android OS. Dalvik Executable (DEX) files, on the other hand, are binary files, and they contain the compiled code. An APK file includes compiled program classes in DEX format.

\begin{figure}[h]
\centering
\includegraphics[width=.35\textwidth]{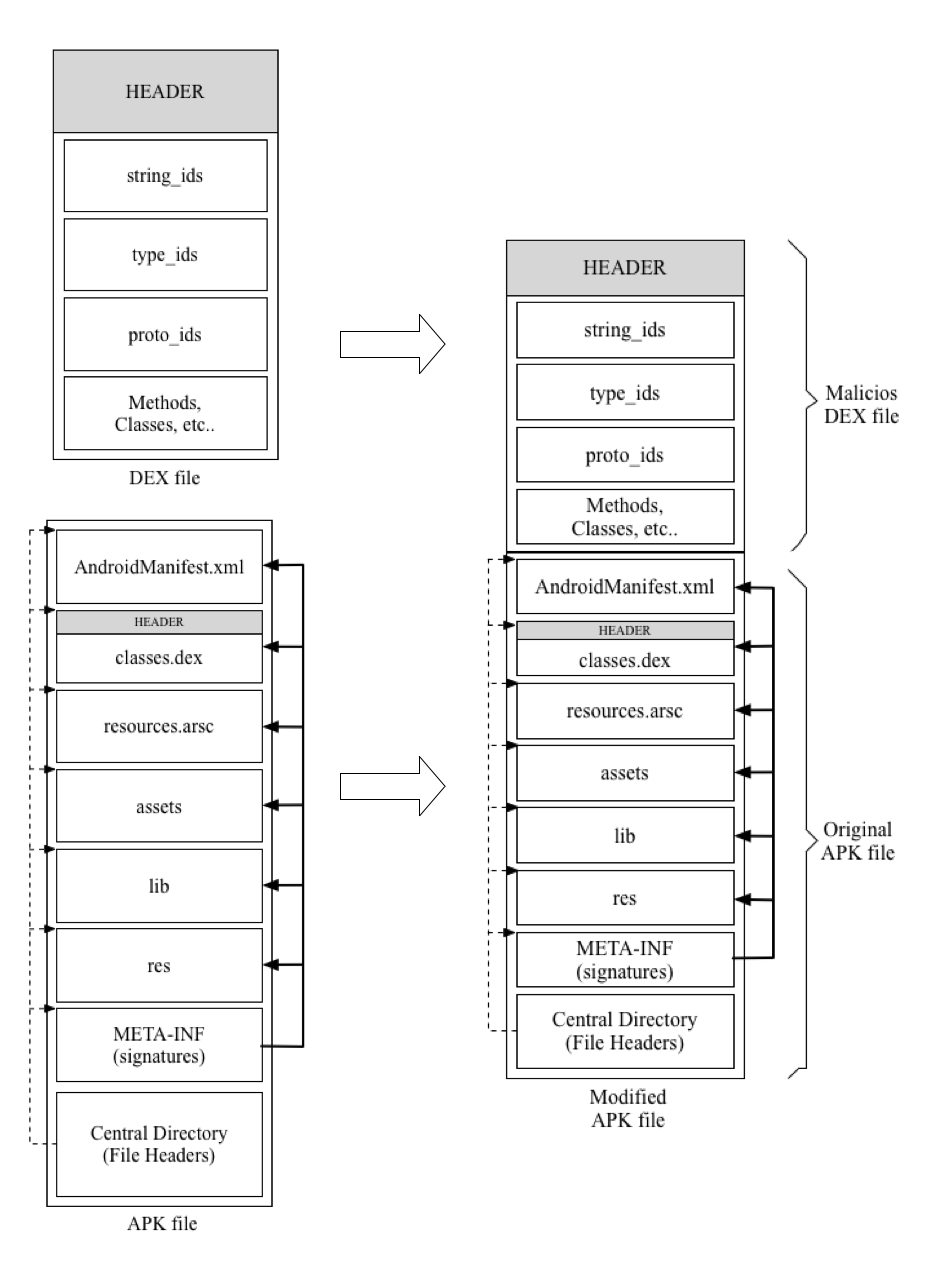}
\caption{Embedding DEX into APK files}
\label{fig:janus}
\end{figure}

A vulnerability called \emph{Janus} in some certain versions of the Android OS allows attackers to create a modified APK file from a legitimate APK and a malicious DEX file without changing the app signature as shown in Figure~\ref{fig:janus}. The modified APK file is then recognized as a legitimate app, and can be installed to the device.

This vulnerability affects apps running on devices with Android 5.0 to Android 7.0 and signed with APK signature scheme v1~\cite{guardsquare2017}. A security firm reported the issue, and Google released a patch in November 2017 to prevent this vulnerability to affect new devices that have Android 7.0 and newer OS. However, older devices and apps that have not been signed with APK signature scheme v2 still remain at risk.

\begin{figure}[h]
\centering
\includegraphics[width=0.30\textwidth]{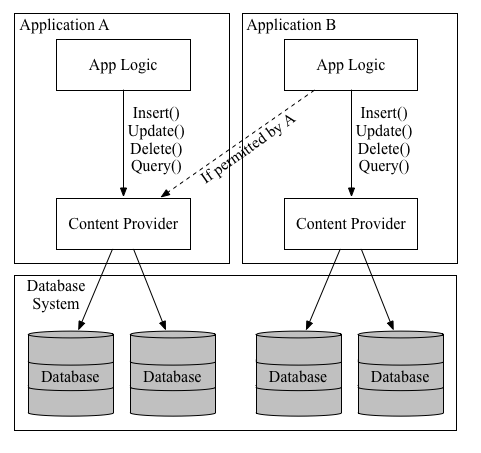}
\caption{Content Provider access}
\label{fig:contentprovider}
\end{figure}

\tinysection{Database Vulnerabilities} Android databases are usually controlled by \textit{content providers}. They are configured in \textit{AndroidManifest.xml} which is a configuration file and is present in every app. Generally, the database of each app is private and do not allow access from other apps. However, it is possible to configure the content provider to permit other apps to query the database as seen in Figure~\ref{fig:contentprovider}. It is also possible to access databases through calling database instance directly from within the app. In this case, the database is definitely unique to the app, and the database cannot be shared with other apps.

A simple configuration error such as unintentionally setting \texttt{exported=true} can open up the content provider to the use of other apps and service requests. ContentScope~\cite{jiang2013detecting} reports that out of 62,519 apps they have surveyed, 2,150 apps had their content providers exposed. Of course, it is possible for some apps to allow access to their content providers. However, it is also possible that the app developers used an example code they found on a website like \textit{stackoverflow.com}. DBDroidScanner~\cite{hassanshahi2017android} identifies further vulnerabilities in the content providers, and alternative ways of creating \emph{SQLite} databases within apps. Furthermore, some of the apps keep data stored in the database in plain text, and subject to synchronization manipulation~\cite{jain2016androiddatabase}.

\tinysection{Attack vector} One of the threats Janus vulnerability poses is embedding a DEX file that accesses the app database that contains sensitive information. If the injected code does not change any other function, and just focuses on stealing data, the app will still function  normally, and the users will not be aware of their information being stolen. However, it is also possible for the injected code to tamper with the database contents, hence affecting what the user does and sees, and changing the app behavior.

The content provider and SQLite interfaces provide access to the database, and makes it possible for the attackers to issue raw SQL queries to the app's database. Again, since the app itself is benign, and has not been changed, it is possible for any user to be subject to these attacks.

\subsection{Threat Model}
\label{sec:threat}

Janus is a vulnerability that originates from how Android validates if an APK file is legitimate. It injects malicious code into the app by adding binary code to the original file. The malicious code runs alongside the app, where it can perform a number of activities such as accessing the database to steal information. Janus is not the first vulnerability that had similar repercussions -- HTML5-based apps had been a target for code injection attacks~\cite{jin2014code} along with other web-to-app attacks~\cite{hassanshahi2015web}. Content provider vulnerabilities, on the other hand, expose directly the data access layer of an app to the attackers, and the DBDroidScanner~\cite{hassanshahi2017android} reports that the number of vulnerabilities and vulnerable apps are growing. 

Our system aims to detect attacks that originate from known and unknown vulnerabilities that expose the sensitive information in the database to attackers.
We target attacks that query the database to glean sensitive information, and tamper database records as a result of vulnerability types presented above.
%We target both passive and active attacks that query the database to glean sensitive information, and tamper database records as a result of vulnerability types presented above. Active and passive attacks are distinguished based on altering system resources or affecting their operation~\cite{stallings2017cryptography}. \textit{Passive attackers} query the database to find sensitive data. \textit{Active attackers}, on the other hand, tamper with the data and query results.
These attacks can behave and affect the workload in three different ways:

\tinysection{Copycat attack model} The workload created by the app stays unaffected, the malicious code creates an additional workload with its own query generation strategy.

\tinysection{Free-styler attack model} The workload created by the app stays unaffected, the malicious code creates an additional workload with the app's query generation strategy.

\tinysection{Translator attack model} The workload created by the app gets affected by malicious code through the overridden classes and information flow changes. This model can be subdivided into two categories: (1) The malicious code modifies the query generation strategy to easily extract information required by the attackers, and (2) The malicious code modifies the information flow which results in the app generating legitimate queries for actions that do not require them.

%Our system does not block such behavior, but it observes the query activity of the user, and raises an alert. \todo{The alert can force the user to update/download the app from the Google Play Store and/or forward the behavior report to the app developers.}

%However, our system does not address \textit{snapshot attacks} in which the adversary copies the database server, or the database instance completely. Such attacks cannot be caught by the query monitor since it doesn't have any control on the server instance.
%The cloned server instance, if the attack is successful, would be available for use of the adversary. In that case, the adversary wouldn't need to go through the query access monitor to query the database.

%This model specifically effective against preventing insider attacks to databases.

\subsection{Architecture}
\label{sec:architecture}

\begin{figure}[h]
\centering
\captionsetup[subfigure]{justification=centering, skip=1pt}
    \centering
            \begin{subfigure}[b]{0.49\textwidth}%{0.49\textwidth}
        \centering
        \includegraphics[width=1\textwidth]{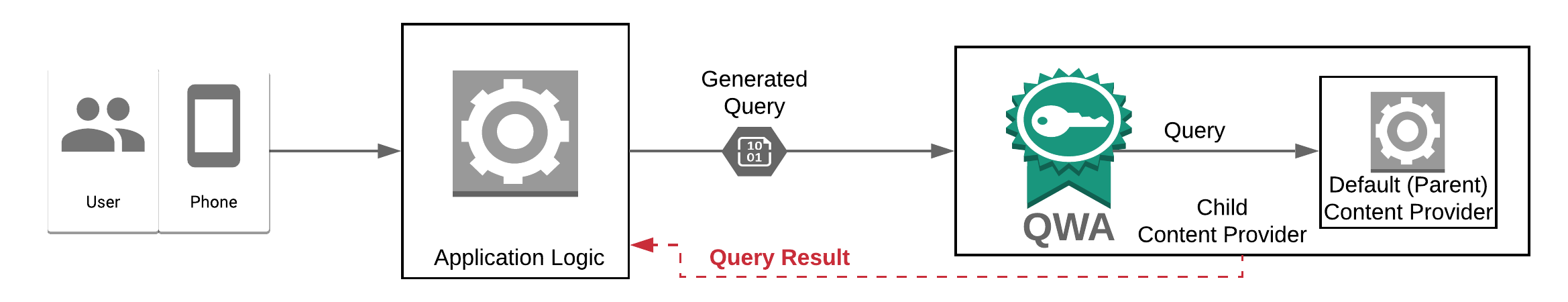}
		\caption{QWA as an extension to the Content Provider}
        \label{fig:overviewExtendedCP}
    \end{subfigure}\\
    \begin{subfigure}[b]{0.49\textwidth}%{0.49\textwidth}
        \centering
        \includegraphics[width=1\textwidth]{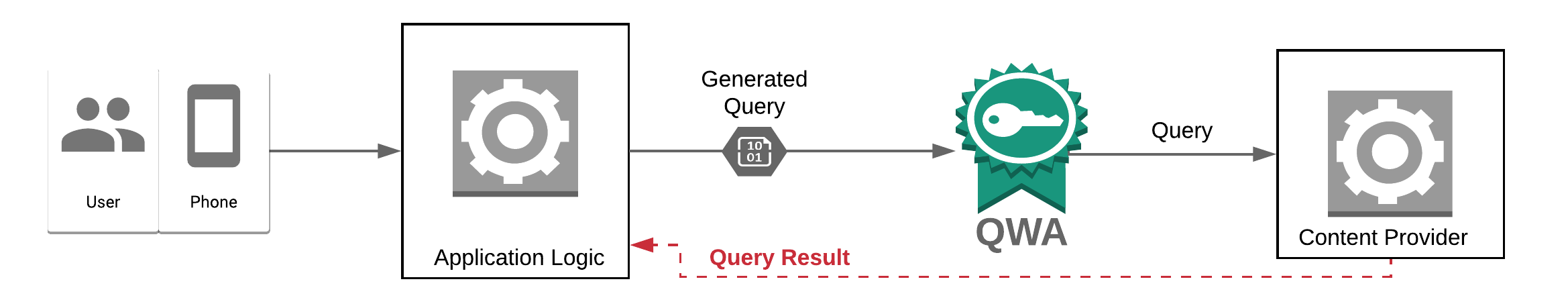}
		\caption{QWA as a Mediator}
        \label{fig:overviewWithoutCP}
    \end{subfigure}
\caption{Architecture}
\label{fig:overview}
\end{figure}

QWA is designed to be modular and flexible, in order to be able to integrate with the other OS security features, apps and various databases.
If it is implemented in the OS layer as an extension to the default \texttt{ContentProvider} as shown in Figure~\ref{fig:overviewExtendedCP}, after it is active and running, any app that interacts with the database can be monitored. 
Also, it can be integrated into any app just to observe the query traffic so that it does not bring timing overhead to the query processing. If it is integrated between the application logic and the \texttt{ContentProvider} as shown in Figure~\ref{fig:overviewWithoutCP}, it acts as a mediator between the app and the \texttt{ContentProvider}.
It is also possible to extend the \texttt{ContentProvider} within the app by inheriting the \texttt{ContentProvider} class, and overriding its methods.
% as illustrated in the code stub given in Figure~\ref{fig:ExtensionCode}.

%\begin{figure}[h]
%\centering
%\includegraphics[width=0.5\textwidth]{graphs/codeSnippet}
%\caption{Extension Code}
%\label{fig:ExtensionCode}
%\end{figure}

When a user uses an app on their device through the app graphical user interface (GUI), the app generates queries, and issues them to the database.
%On the other hand, the terminal users interact with the database via a terminal interface on their computers, which directly connects them to the database server.
The database is contained in a database server instance running on the device.
QWA just observes the queries that are issued to the database, and it does not block or change the queries.
Any query that is issued to the database is captured by the QWA, processed, and logged there, and then sent to the database.
Although QWA does not block any queries, it detects suspicious activity, and reports them.
The overview of the system architecture is depicted in Figure~\ref{fig:overview} where QWA acts as the observer in the system.

\section{Methodology}
\label{sec:usermodeling}
% !TEX root = ../paper.tex

In this section, we describe
%the threat model used in this research, 
our user behavior modeling methodology, and the anomaly detection strategy.

\subsection{SQL Query Feature Extraction}
\label{sec:featureextraction}

A \textit{relational database} is a set of relations (\textit{i.e., tables}), and a \textit{relation} is a bag
%or set
of tuples (\textit{i.e., rows}) where a tuple is a structured data item. The structure is defined with attributes (\textit{i.e., columns}) and the types of the attributes.
%The difference between a bag and a set is that bag semantics allow duplicate records whereas set semantics enforce uniqueness.

%Relational Database Management Systems (RDMS) ensure result accuracy while providing efficiency for joining combination of multiple resources.
%The most important principle in these systems is that a query simply must not return a wrong result.

\textit{Structured Query Language (SQL)} is a declarative language that is designed for managing, manipulating, and retrieving from relational databases.
Other than schema manipulation and data access control operations, SQL queries mainly perform 4 different operations: (1) insert, (2) update, (3) delete, and (4) select. The basic structure of these operations are given respectively:

{\footnotesize
\begin{verbatim}
(1) INSERT INTO table (column1, column2, ...)
               VALUES (value1, value2, ...);
(2) UPDATE table
    SET column1 = integer|decimal|string|...
    WHERE column2 = integer|decimal|string|...
(3) DELETE FROM table
    WHERE column1 = integer|decimal|string|...
(4) SELECT [aggregation] column1, column2, ...
    FROM table1, table2
    [WHERE table1.column1 = table2.column3]
    [ORDER BY column1] [GROUP BY column1]
    [LIMIT integer]
\end{verbatim}
}

where the brackets show optional query items. As can be implied from these basic structures, queries that perform similar tasks usually have analogous structures, or at least share some attributes. 
SQL query statements are constructed from \textit{clauses}. Every line of the query structures given above constitutes a clause. As an example let's take the following query which reads as \textit{``Show the names and number of games played for each player who is over 30''}:

{\footnotesize
\begin{verbatim}
(1) SELECT p.name, COUNT(g.played)
(2) FROM player p, game g
(3) WHERE p.id = g.playerid AND p.age > 30
(4) GROUP BY u.name
(5) ORDER BY u.name
\end{verbatim}
}

Line 1 consists of the \texttt{SELECT} keyword, and the \textit{projection} items. Line 2 has the \texttt{FROM} clause which lists the tables the query is going to use. Line 3 is named the \texttt{WHERE} clause. \texttt{WHERE} clause contains \textit{selection} and \textit{join} expressions. \texttt{p.id = g.playerid} expression is a join expression, and \texttt{p.age > 30} is a selection expression. Line 4 and 5 include the \textit{group-by} and \textit{order-by} items, respectively.

%Auditing SQL query logs of databases is being used for various purposes from security~\cite{Sun2016} to database performance optimization~\cite{Bruno:2005:APD:1066157.1066184}. The query logs generally include usernames, timestamps, and other parameters along with the query, hence letting the auditor to collect useful information. However,
Query interpretation, namely, understanding the goal of the query, is regarded as hard as creating a new query, and even more so for complex queries~\cite{gatterbauer2011databases}.
Furthermore, complex queries are not uncommon due to the expressive nature of the SQL. The databases are designed and optimized for performance and correctness, which requires simple relations.  Hence, the queries need to be designed more complex with high numbers of table joins as the need increases to access complex information.
There is a line of research that aims to capture user intention through queries since it would contribute to security applications~\cite{kul2016ettu}, automated personalized query generation~\cite{magda2010snipsuggest}, and interest mining~\cite{stefanidis2009you}. 
To accomplish this, it is essential to identify the required features to be extracted from the SQL queries.
As mentioned before, the data stored in the database can also be a good indicator for measuring the similarity of the queries~\cite{Mathew2010Raid}. For instance, consider the following query:

{\footnotesize
\begin{verbatim}
	(1) SELECT * FROM contact
	    WHERE name LIKE "A%"
	(2) SELECT * FROM contact
\end{verbatim}
}

The first query reads as \textit{``list all contacts whose names start with A''}, and the second query reads as \textit{``list everything on contact table''}. 
However, if this query runs on a table where there is only one contact whose name is \textit{Alice}. Thus,
SQL queries are also open to varying \textit{interpretations}. Consequently, it is crucial to have a SQL query extraction strategy according to why these features are required. For instance, query recommendation requires analysis of feature correlation and dependency~\cite{magda2010snipsuggest}, while performance optimization requires discovery of table joins~\cite{aligon2014similarity}.

As discussed before, we observe and process every SQL query issued to the DBMS. We extract the relevant features of the query considering which part of the database the user is accessing with that particular query. In our previos work~\cite{kul2018tkde}, we investigated the query clustering quality of several query feature extraction methods. Our work follows the basic principles of the two most commonly used SQL query feature extraction methods~\cite{aligon2014similarity, makiyama2015text}. 

%Aouiche \textit{et al.}~\cite{aouiche2006} aim to optimize view selection in data warehouses by the queries posed to the system.
%They consider the \textit{selection}, \textit{join} and \textit{group-by} items in the query to create vectors and use Hamming Distance to measure how similar two queries are.
%While creating the vector, it doesn't matter if an item appears more than once or where the item is.
%They cluster similar queries that create a workload on the system and base their view creation strategy in the system on the clustering result. 

%Kamra \textit{et al.}~\cite{Kamra2007SyntaxBased} use database query logs to detect insider attacks. They construct a \textit{quiplet} out of SQL queries, which essentially acts as a vector. The quiplet has 5 different substructures: \textit{command}, a set of tables used in \textit{projection}, projected attributes, a set of tables used in \textit{selection} and \textit{join}, and attributes used in the selection and join. To be able to construct a quiplet when a new query arrives, it is required to possess the database schema structure.

Aligon \textit{et al.}~\cite{aligon2014similarity} survey on comparing OLAP sessions considering the query similarity, and session similarity.
They classify \textit{selection} and \textit{join} attributes as the most relevant component in a query followed by the \textit{group-by} attributes.
With the light of the findings, they propose their own SQL query extraction schema which considers \textit{projection}, \textit{group-by}, \textit{selection} and \textit{join} attributes for queries issued on OLAP datacubes. Makiyama \textit{et al.}~\cite{makiyama2015text} focus on workload exploration on large query logs.
They extract the attributes in \textit{selection}, \textit{join}, \textit{projection}, \textit{from}, \textit{group-by} and \textit{order-by} items separately, and record their appearance frequency. 
%They create a feature vector using the frequency of these terms which they use to calculate the pairwise similarity of queries with cosine similarity.
%Instead of clustering, they perform the workload analysis with Self-Organizing Maps (SOM).
We approach query feature extraction with the goal of understanding which part of the database the query writer is interested in.
We extract the terms in \textit{selection}, \textit{joins}, \textit{projection}, \textit{group-by}, and \textit{order-by} items along with \textit{constants} and \textit{parameters} in the query separately, and record their appearance frequency.

\subsection{Normal Behavior}

Users access information on the database 
%either through a program, or through a SQL querying interface like MySQL Workbench or a terminal screen, which are then processed in the organization's database servers.
%Autonomous systems, on the other hand, usually issue the preprogrammed data access requests directly to the organization's database servers.
through interacting with the app. The app generates queries based on the activities performed, and retrieve data from the database with these queries.

Building user profiles through clustering, and other machine learning techniques has been studied extensively in the literature before~\cite{Kamra2007SyntaxBased, Mathew2010Raid, wang2010hengha}. However, this approach is not suitable to make the user profiles adapt to the behavior changes, or to allow the anomaly detection strategy to consider a possible behavior shift. They usually take a snapshot of all the activity at a certain time, and create a model based on the information available at that point of time without even considering the activity time. Since the query set is clustered with an uncertain number of labels, it is required to compute a pairwise distance matrix between queries to perform a clustering with hierarchical clustering or a similar clustering method. This operation has quadratic complexity~\cite{chandola2009anomaly}, and is required to be performed over the whole set of queries. When the model starts to perform worse, the re-training of the model requires the same operation to be repeated.

We focus on observing behavior in individual profiles to show the importance of behavior drift. For each user, we define a user profile for a given timeframe $T$, denoted as the vector $\phi_u^T \in \mathbb{R}^d$, where $d$ is the total number of features extracted using the methodology given in Section~\ref{sec:featureextraction}. To compute $\phi_u^T$, we consider all queries issues by the user $u$ to the database within the timeframe $T$. A query issued at time $t$, is a $d$ length vector of counts, and is denoted as $q_u^t$, where the $i^{th}$ element, $q_u^t[i]$, is equal to the number of times the feature $i$ is observed in the corresponding query. 
 %
% We collect these queries issued to the database along with supporting information such as usernames, and timestamps in order to create user profiles. A query, issued by user $u$, at time $t$, is denoted as a $d$ length vector, $Q_u^t$, where $d$ is the total number of features extracted using the methodology given in Section~\ref{sec:featureextraction}, and denoted as:
%
%\begin{equation}
%Q^{t}_u = ( f^{t}_1(c_0), f_2^{t}(c_1), ... , f_m^{t}(c_n) )
%\end{equation}
%
%where $t$ is the timestamp that the query $Q$ was issued, $u$ represents the username of the query owner, $f_i$ is the feature extracted, and $c_j$ denotes how many times the feature $f_i$ was observed in the query.

Note that the feature extraction from a query is an $O(d)$ time complexity operation where $d$ is a relatively small number, compared to the number of queries.% in computational standards, since we process all the features of a query. Even if a query has hundreds of features, this operation is computationally very small.

By combining the feature counts across all queries issued by the user $u$ in a given timeframe, one can compute the entries in the user profile vector, $\phi_u^T$, as follows:
\begin{equation}
\phi_u^T[i] = \frac{\sum_{\forall t \in T}{q_u^t[i]}}{\sum_{j=1}^d{\sum_{\forall t \in T} {q_u^t[j]}}}
\end{equation}
%User activity $A$ is represented by a user $u \in U$, where $U$ is the set of all users, for the time period $T$ that starts from $t_0$ and goes on for $\Delta t$, and the set of queries $Q$ performed by $u$ within $T$. Formally,
%
%\begin{equation}
%A^T_u = ( Q^{t_0}_u, Q^{t_1}_u, ... , Q^{t_n}_u )
%\end{equation}
%
%where $Q^{t_i}_u$ represents a query $Q_{t_i}$ issued at time $t_i$ by user $u$.

The \textit{user profiles} are created with the accumulation of these features for a given period of time.
Using the appearance frequency of these features, we calculate the appearance probability of each harvested feature. One can also consider the user profile for timeframe $T$, as a multinomial random variable, which can take one out of $d$ possible values, with probability distribution parameterized by $\phi_u^T$.
%This multinomial probability distribution of the features for each user constitutes the \textit{user profile}.
%When a feature is no longer in the time window specified, it is marked as expired, and removed from the user profile.
%A user profile $\phi$ is formally denoted as:
%
%\begin{equation}
%\phi^T_u = ( P(f_0)^{T}_u, P(f_1)^{T}_u, ... , P(f_n)^{T}_u )
%\end{equation}
%
%where $P(f_i)^{T}_u$ represents the probability of encountering feature $f_i$ within the timeframe $T$ among all the operations performed by user $u$.

Given that the features are stored in a map structure, the features of a new query can be simply added to the feature counters which are used to compute the probability of a feature. Hence, this operation has only $O(1)$ time complexity.

Logically, we expect the preprogrammed queries to be more consistent between each other, while handwritten queries to form a more diverse distribution.
For instance, DBAs and data analysts access a variety of data as required by their jobs. However, apps generate queries based on the data access layer's query generation strategy with parameters provided by the methods that use the data access layer. Sometimes, queries can even be hardcoded into the app source code. Therefore, query diversity is expected to be lower than handwritten queries.
%Even for handwritten queries, some roles require adaptation to new tasks which would increase the diversity of the features in the user profile.
%For instance, DBAs and data analysts access a variety of data as required by their jobs. 
%However, an HR intern whose job is only retrieving the applications to the company and reporting them to their superior, diversity is expected to be very low.
As a result, 
we define this expected change in behavior with the term \textit{profile drift}.

%We compute the difference between distributions with Kullback--Leibler Divergence~\cite{kullback1951information}. 
Comparison of the accumulated user profile, for timespan $T_1$, with the new incoming behavior observed for timespan $T_2$, using Kullback--Leibler Divergence~\cite{kullback1951information} gives the \textit{drift score} denoted as follows:

\begin{equation}
d^{T_2}_u (\phi^{T_2}_u || \phi^{T_1}_u) = \sum_i \phi^{T_1}_u(i)  log_2 \frac{\phi^{T_1}_u(i)}{\phi^{T_2}_u(i)}
\end{equation}

\textbf{KL-Divergence (\textit{i.e., relative entropy})} is used for comparing two probability distributions, $P$ and $Q$; and it ranges between 0 and $\infty$. $D_{KL}(P||Q)$ essentially represents the information loss when $Q$ distribution is used to approximate $P$.

Note that when $P(i) \neq 0$ and $Q(i) = 0$,  $D_{KL}(P||Q)=\infty$. For example, suppose, we have two distributions $P$ and $Q$ as follows: $P = \{ f_0: 3/10, f_1: 4/10, f_2: 2/10, f_3: 1/10 \}$ and $Q = \{ f_0: 3/10, f_1: 3/10, f_2: 3/10, f_4: 1/10 \}$. In this case, since $f_3$ is not a part of $Q$, the result would be $\infty$, which means these two distributions are completely different. 

\textbf{Smoothing.} To get past this problem, we can apply \textit{smoothing} (i.e., Laplace/additive smoothing), which is essentially adding a small constant $epsilon$ to the distribution, to handle zero values, without significantly impacting the distribution. After we apply smoothing, $D_KL(P||Q)$ becomes $1.38$.

The intuition behind using KL-Divergence in our method is to identify the change we experience in the newly coming behavior, compared to the base profile. Similarly, the intuition behind using smoothing is to assume that even if a feature has not been seen in the given dataset, we can still take into account the possibility of its appearance, although very small. Without smoothing, distributions with thousands of matching features except one could be regarded as not related.

%\todo{Discuss the time complexity of the operation.}

%The similarity between user profiles, and the difference between distributions are computed with Kullback--Leibler Divergence~\cite{kullback1951information}.

\subsection{Anomalous Behavior}

The user profile evaluates as the new features from the newly coming behavior are imported to the profile. However, before they are added to the profile, they are tested to see if this new activity is an anomalous behavior.

The drift scores over time, which form a vector denoted as $DS$, are used to calculate the linear regression coefficients to see the ordinary behavior change for the user. The resulting model function of linear regression of these drift scores for a given period of time yields the \textit{profile drift}, formulated as follows:

\begin{equation}
\hat{DS}_i = \hat{\beta}_0 + \hat{\beta}_1 t_i + \hat{\epsilon}_i
\end{equation}

 where $\hat{\beta}_0$ is the y-axis intercept constant, and $\hat{\beta}_1$ is the slope of the profile drift line. $\hat{\epsilon}_i$ is a very small number that represents the noise.

To compute $\hat{\beta}$, given a matrix $C \times N$, the naive least squares computation has overall $O(C^2 N)$ complexity, or we can use LU or Cholesky decomposition which takes $O(C^3)$ where $C$ denotes the number of features and $N$ denotes the number of training examples. Since we can usually assume $N > C$, $O(C^2 N)$ dominates $O(C^3)$. As a result we can consider that the total asymptotic complexity for linear regression is $O(C^2 N)$. However, since we are using the KL-Divergence score of two probability distributions, the number of features scales down to 1, which results in the complexity of this step scaling down to $O(C^2)$ where $C$ is expected to be a very low number by computational standards. For instance, if we take the profile drift computation interval as one day, we end up with $C = 365$ for a year of data.

Profile drifts occur as the users take on different tasks, as they start to grow different interests, or as they gain experience on the job.
Consequently, when this constant change is not addressed properly, utilizing a predefined threshold value can lead to raising too many false positives for the security personnel to inspect when it is set too low, or too many false negatives when it is set too high to avoid false positives.
Hence, we define an anomaly as a drift score larger than the sum of the expected profile drift at that specific point, and expected error (\textit{i.e., standard deviation}) as follows:

\begin{equation}
    func(\phi^{T_2}_u)= 
\begin{cases}
    raise alarm, & \text{if } d^{T_2}_u > \hat{DS}_i + \sigma_{DS} \\
    normal behavior, & \text{otherwise}
\end{cases}
\end{equation}

Positive drift in a profile implies that the user is inclined to change their behavior rapidly. Negative drift at any point intuitively suggests that the user started not to get out of their usual pattern as much as before. Issuing no queries at all does not cause any security concerns while increasing the sensitivity to behavior drift when the user starts to issue queries again. When a system uses our model, by using a sliding window strategy, this high sensitivity will fade away as the behavior drift will converge in time.

\section{Experiments}
\label{sec:experiments}
% !TEX root = ../paper.tex

In this section, we first describe our experimental setup and the dataset. We then show how behavior drift can identify different users. We finally present findings from the evaluation of our framework with a real-world SQL query workload.

\subsection{Experimental Setup}

In our experiments, the tests were performed on macOS High Sierra v11.0.1 on a 2.7 GHz Intel Core i5 with 8GB memory.
All of the implementations are performed with Java 1.8 and Python 3.5.

\subsection{Dataset}

%The dataset properties we need for this research are threefold: (1) users should use an application to interact with the database, (2) there should be different tasks that the users can perform, and (3) there should be multiple users using multiple applications to be able to investigate if our findings hold for different settings.

We use Android smartphone query SQL query logs in our experiments. The experiment dataset consists of SQL logs that capture all database activities of 11 Android phones for a period of one month. SQL queries collected are anonymized, and some of the identified query constraints are deleted for IRB compliance. %~\cite{pocketdata}.
In this dataset, the queries are generated by the Android applications. There are 45,090,798 queries in total in this dataset.
%Since some of the constants are replaced with standard placeholders for IRB compliance, the number of distinct queries drops significantly. 
%With the help of the operating system, some Android applications can block query monitoring with privacy concerns such as banking applications.
%To be able to use all the queries issued by the applications, we selected several of them that do not interfere with the query monitor, and also produced the largest number of queries among all other applications:
We selected three apps that have the largest volume of database interactions for our experiments:
(1) Facebook,
(2) Google+,
and (3) Google Play.

%Hence, the smartphone query logs satisfy all of the defined requirements by providing insights to the behavior of 11 users on 5 different Android applications with varying query generation strategies over a period of a month.
We also performed the same experiments on smaller, less used applications, that reflected similar results~\footnote{Disclaimer: We do not claim that the apps selected for the experiments have any vulnerabilities that is presented in this work. This does not mean they do not have similar vulnerabilities that can cause data leakage.}. The total query numbers for each application can be seen in Table~\ref{tab:dataset}.

\begin{table}[]
\centering
\caption{Dataset}
\label{tab:dataset}
\begin{tabular}{|c|c|}
\hline
\textbf{Application}                                           & \textbf{\# of queries} \\ \hline
Complete Dataset     & 45,090,798             \\ \hline
Facebook                                                       & 1,272,779              \\
\hline
Google+                                                        & 2,040,793              \\ \hline
%Hangouts                                                       & 974,349                \\
Google Play & 14,813,949             \\ \hline
%Media Storage                                                  & 13,592,982  \\ \hline          
\end{tabular}
\end{table}
%, and (6) all available applications in the dataset together.

%\todo{We can leave a note for the reviewers here saying that we can increase the number of applications anytime if the reviewers would like to see more results. Limiting the number of applications is just to save space.}

Not all queries issued by Android apps are legitimate SQL -- there can be stored procedure calls, and environment variable checks.
The SQL query logs of Facebook, Google+ and Google Play we used for our experiments are  extracted from PocketData dataset~\cite{pocketdata} and available online~\footnote{https://phone-lab.org/experiment/request/}.
%We ignore queries that cannot be successfully parsed by our open-source SQL parser~\endnote{The link is anonymized for submission}.
%We also released the source code we used in the experiments~\footnote{The link is anonymized for submission}.

\subsection{User Similarity}

This experiment aims to show that even though the app generates the queries with the same logic, user behavior affects the distribution of queries.
%Hence, it is possible to distinguish users from each other.
We investigate how similar the users profiles are to each other in this experiment.

\begin{figure}[h!]
	\captionsetup[subfigure]{justification=centering,skip=1pt}
    \centering
    \begin{subfigure}[b]{0.23\textwidth}%{0.49\textwidth}
        \centering
        \includegraphics[width=\textwidth]{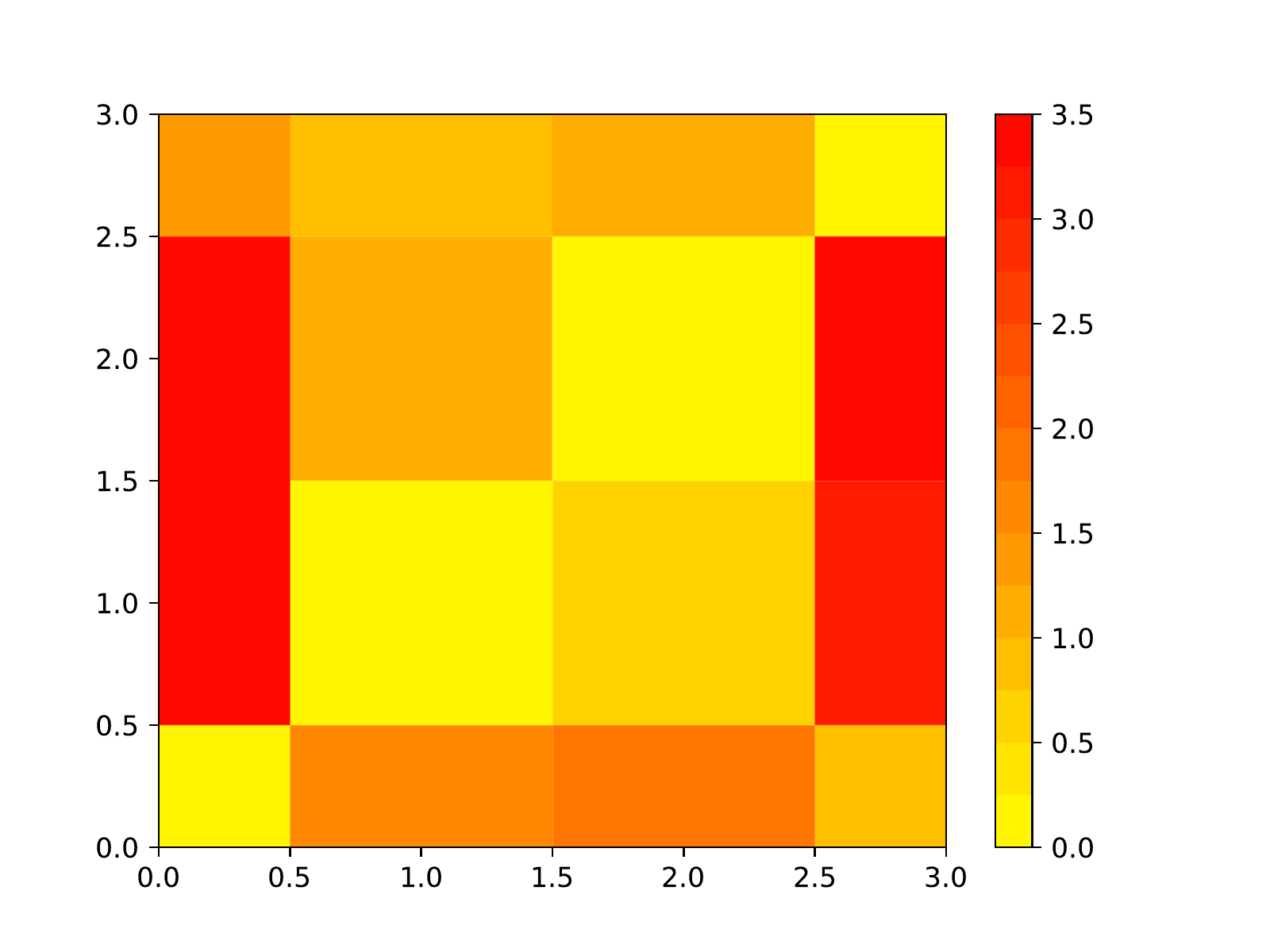}
		\caption{Facebook}
        \label{fig:userDifference:Facebook}
    \end{subfigure}
    ~
    \begin{subfigure}[b]{0.23\textwidth}%{0.49\textwidth}
        \centering
        \includegraphics[width=\textwidth]{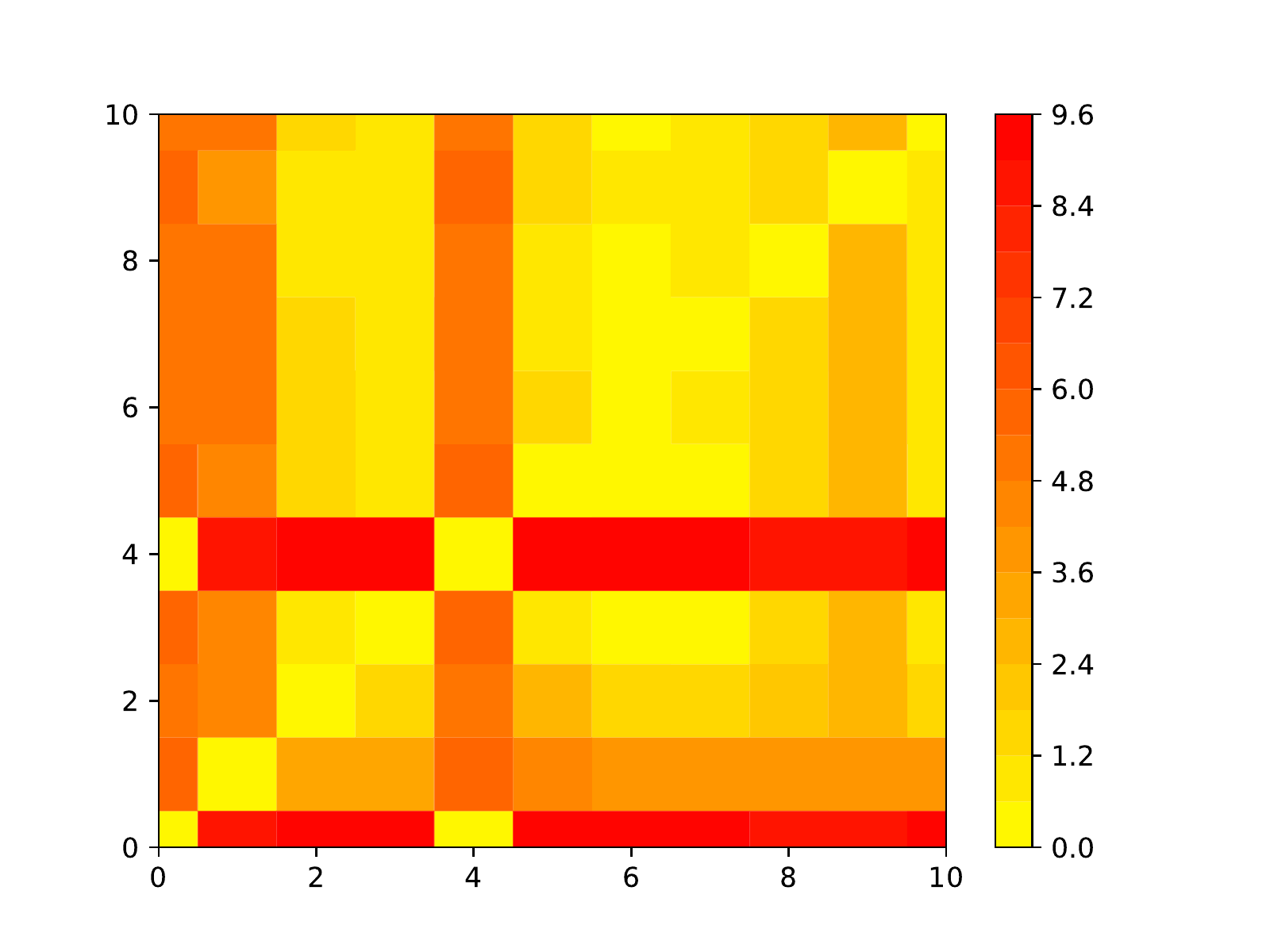}
        \caption{Google+}
        \label{fig:userDifference:GooglePlus}
    \end{subfigure}
    ~
%    \begin{subfigure}[b]{0.32\textwidth}%{0.49\textwidth}
%        \centering
%        \includegraphics[width=\textwidth]{graphs/HangoutsUserDifference}
%        \caption{Hangouts}
%        \label{fig:userDifference:Hangouts}
%    \end{subfigure}
%    \\
    \begin{subfigure}[b]{0.23\textwidth}%{0.49\textwidth}
        \centering
        \includegraphics[width=\textwidth]{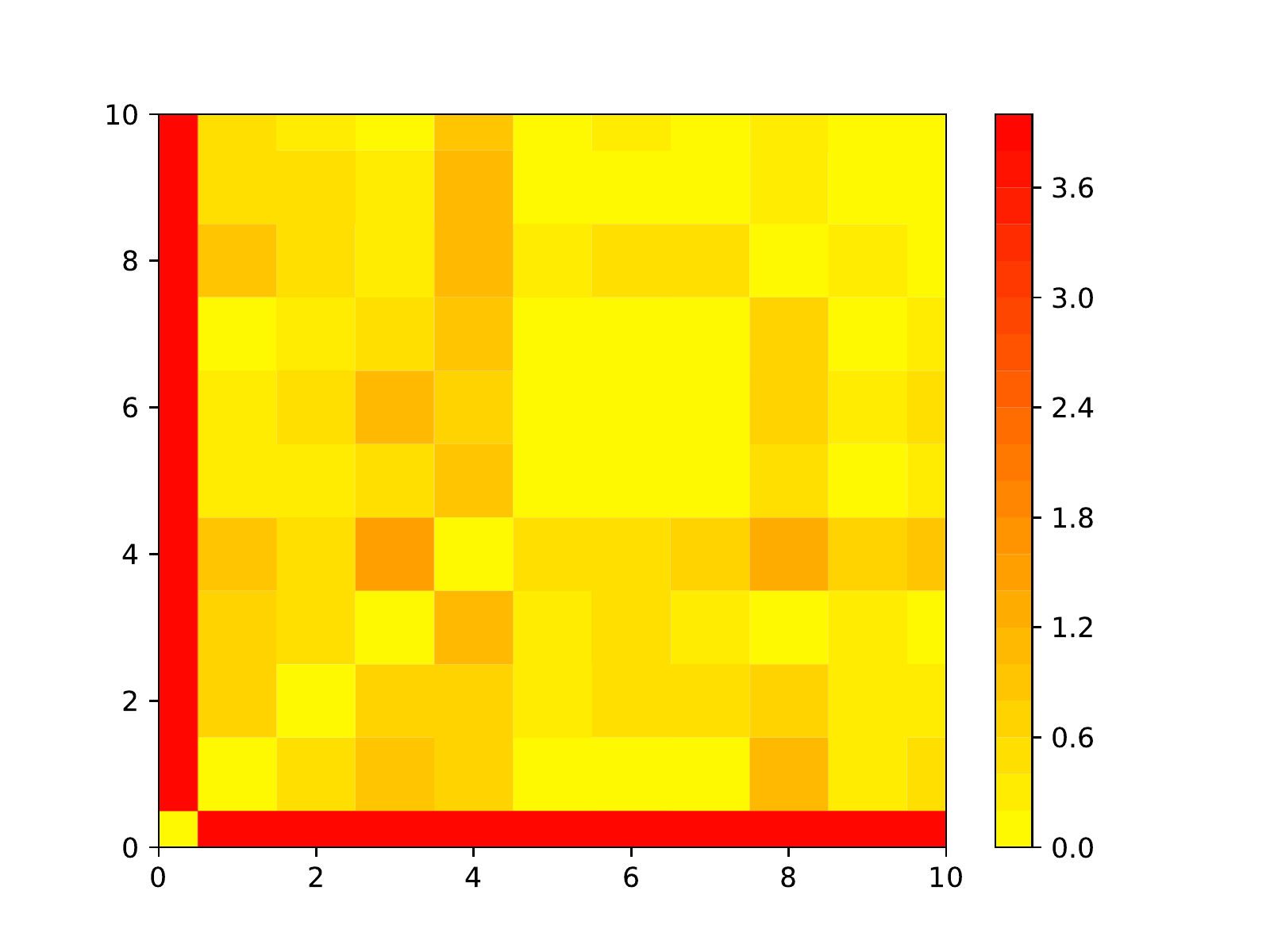}
        \caption{Google Play}
        \label{fig:userDifference:GooglePlayServices}
    \end{subfigure}
%    ~
%    \begin{subfigure}[b]{0.32\textwidth}%{0.49\textwidth}
%        \centering
%        \includegraphics[width=\textwidth]{graphs/MediaStorageUserDifference}
%        \caption{Media Storage}
%        \label{fig:userDifference:MediaStorage}
%    \end{subfigure}
%    ~
%    \begin{subfigure}[b]{0.23\textwidth}%{0.49\textwidth}
%        \centering
%        \includegraphics[width=\textwidth]{graphs/MediaStorageUserDifference}
%        \caption{Complete Workload}
%        \label{fig:userDifference:CompleteWorkload}
%    \end{subfigure}
    \caption{Intra-Application user behavior difference for different Android applications}
    \label{fig:userDifference}
\end{figure}

Figure~\ref{fig:userDifference} shows how comparable different users are with regards to their information access characteristics. In the graphs, darker colors represent that the corresponding user behavior is more distinct while lighter colors represents that the users have similar behaviors. Also note that the color scale is different for each application. We observe that  usage characteristics of Google Play Services are less diverse between users, whereas social media application usage characteristics of users are very distinguishable. Based on Figure~\ref{fig:userDifference}, although the workloads created by different users share the same queries, the distribution of the queries are different. Hence, we conclude that our method can distinguish workloads created by different users.

\subsection{Per-User Behavior Model}

In this section, we report our findings on employing our approach of creating user behavior models for each app.

\begin{figure}[h!]
	\captionsetup[subfigure]{justification=centering}
    \centering
	\begin{subfigure}[b]{0.49\textwidth}%{0.49\textwidth}
        \centering
        \includegraphics[width=0.49\textwidth]{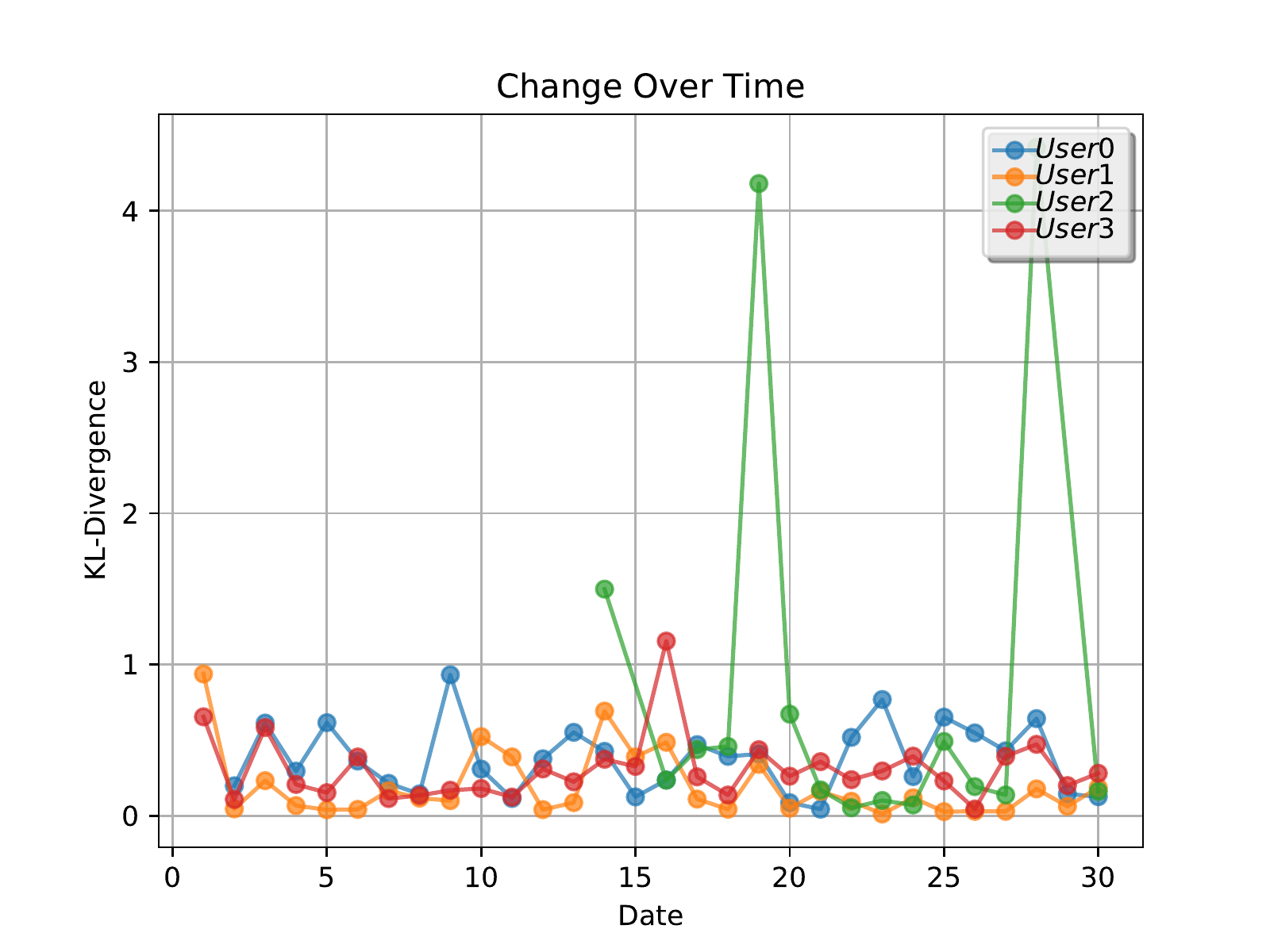}
		\hfill
        \includegraphics[width=0.49\textwidth]{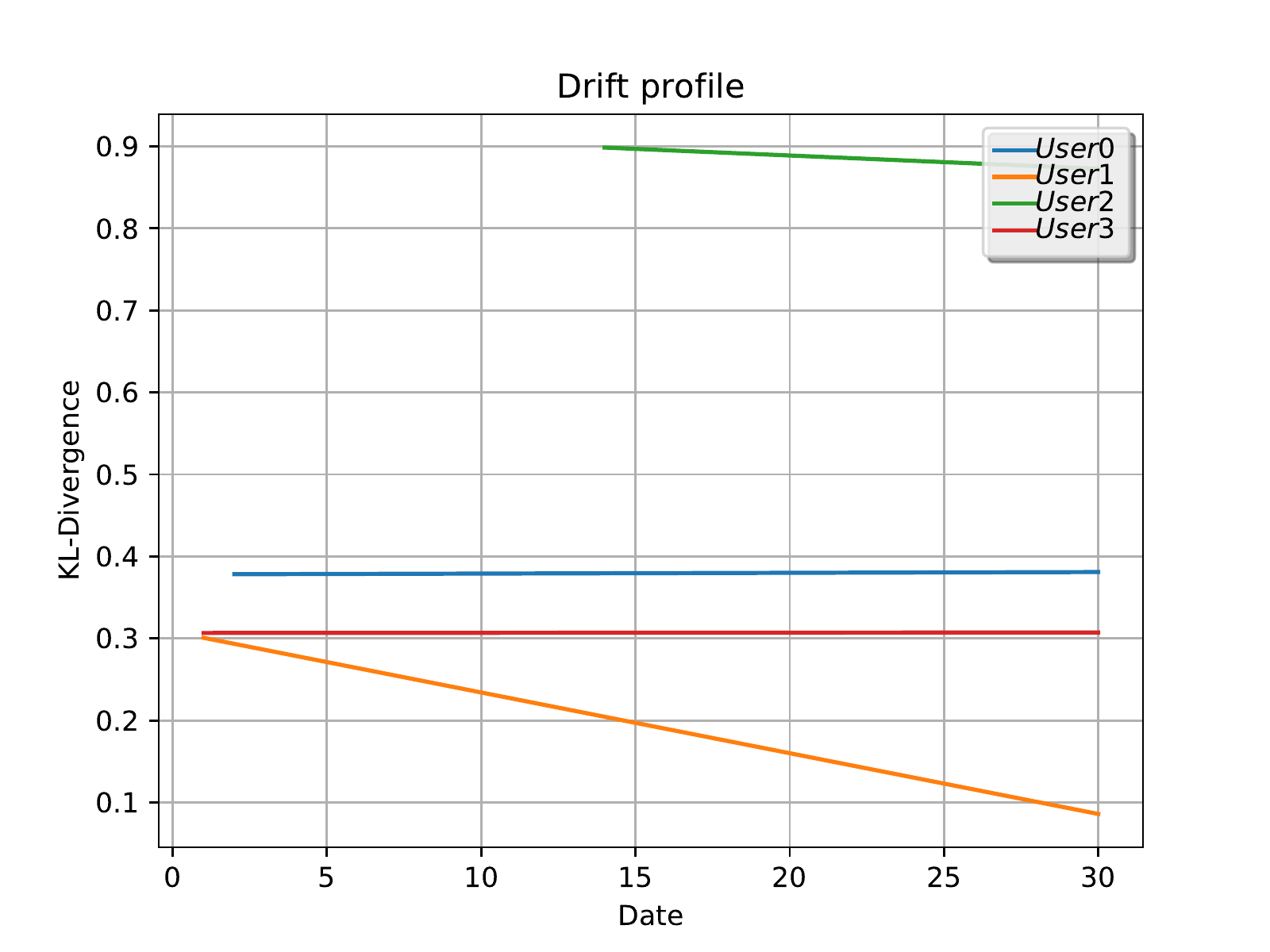}
		\caption{Facebook}
        \label{fig:ChangeOverTimeDaily:Facebook}
    \end{subfigure}
    \\
	\begin{subfigure}[b]{0.49\textwidth}%{0.49\textwidth}
        \centering
        \includegraphics[width=0.49\textwidth]{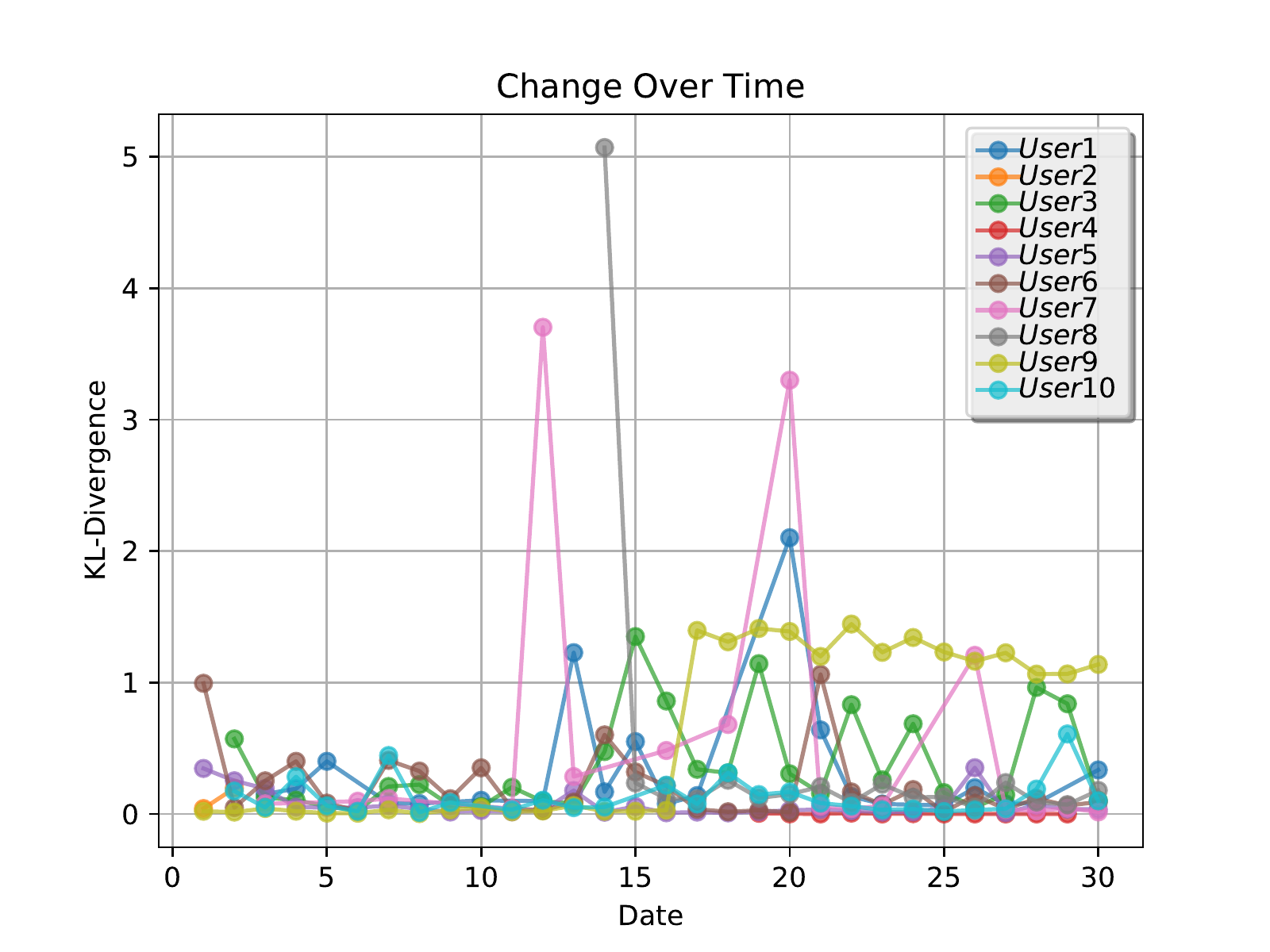}
        \hfill
        \includegraphics[width=0.49\textwidth]{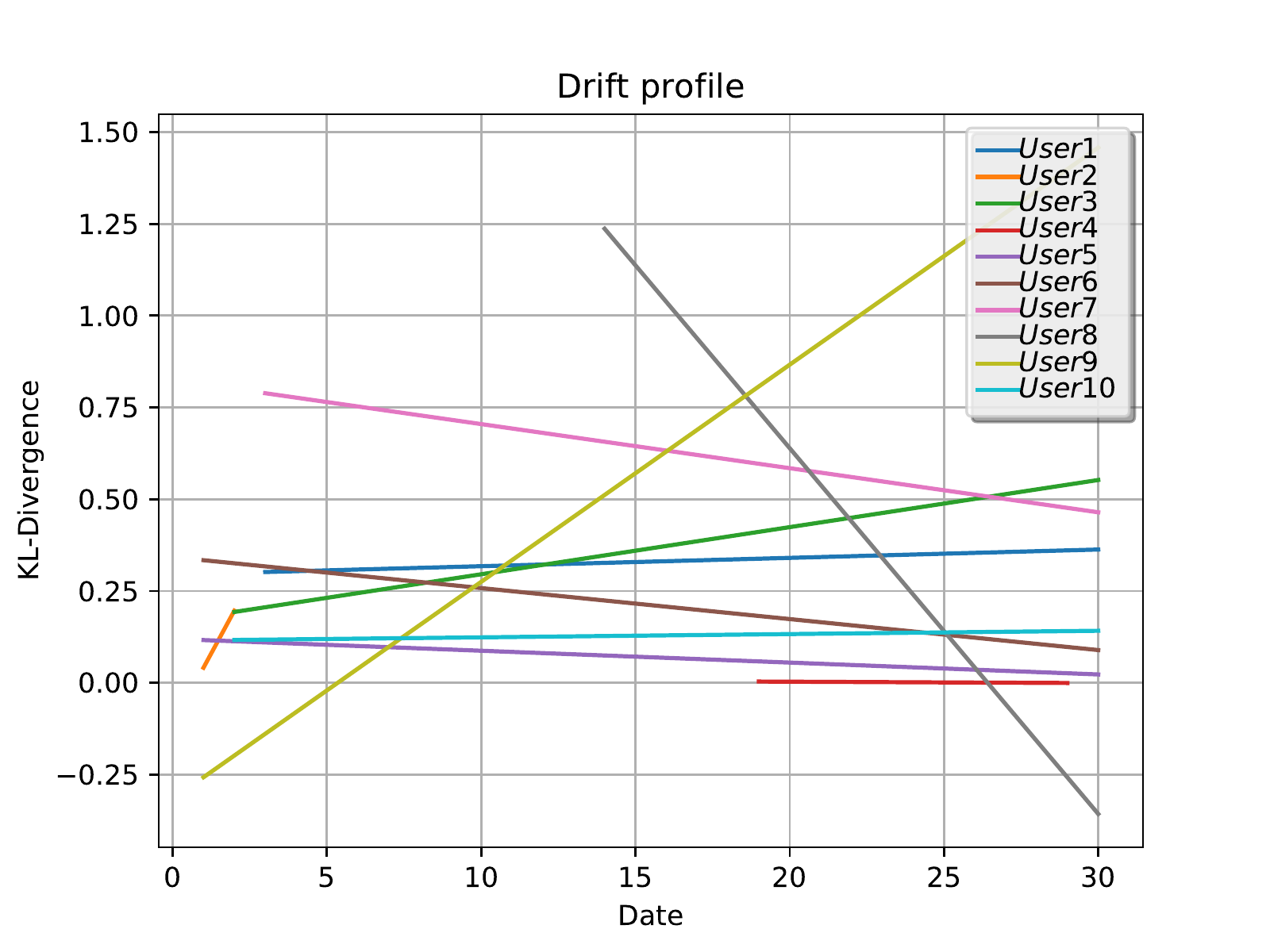}
        \caption{Google+}
        \label{fig:ChangeOverTimeDaily:GooglePlus}
    \end{subfigure}
    \\
    \begin{subfigure}[b]{0.49\textwidth}%{0.49\textwidth}
        \centering
		\includegraphics[width=0.49\textwidth]{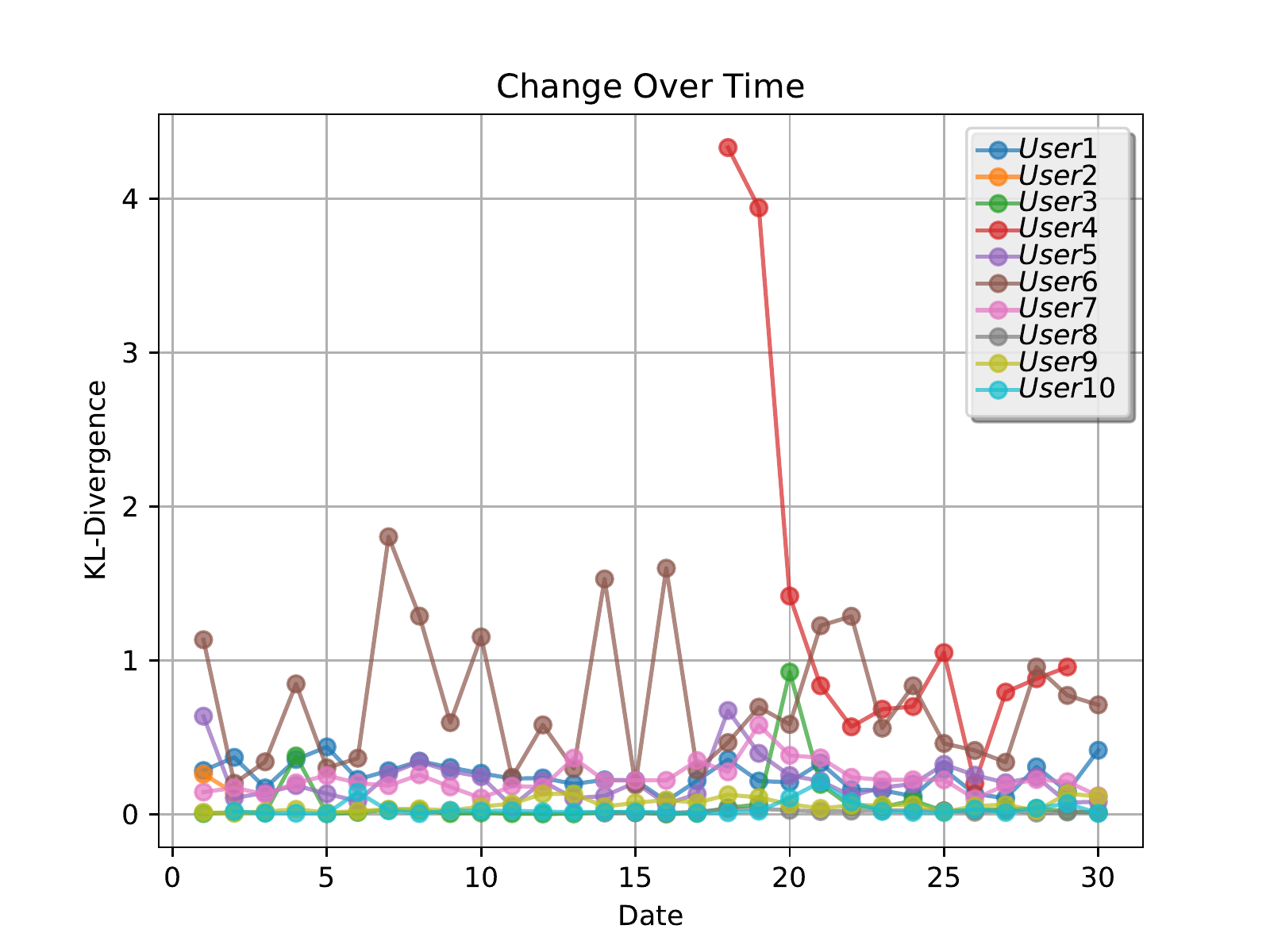}
        \hfill
        \includegraphics[width=0.49\textwidth]{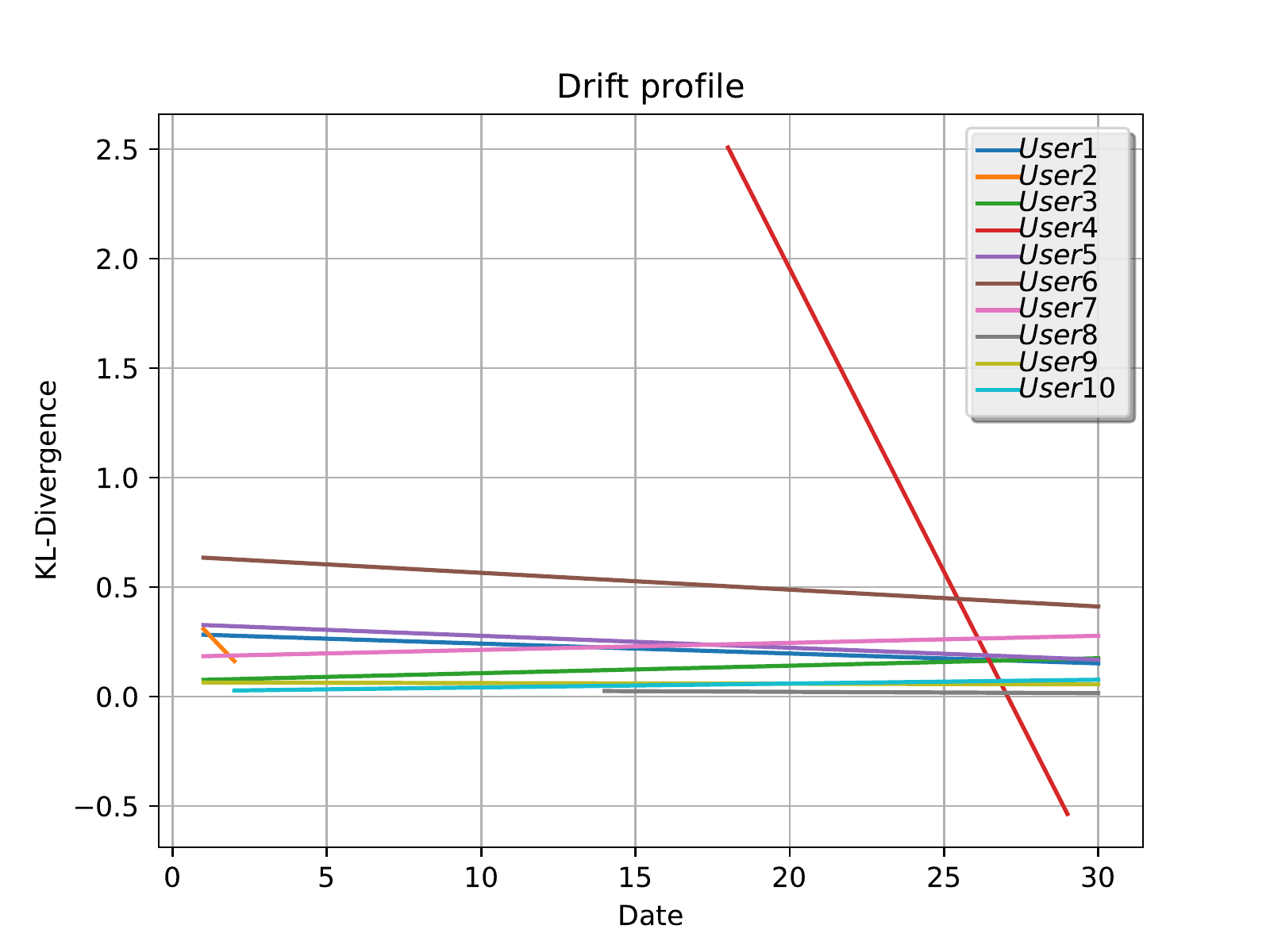}
        \caption{Google Play}
        \label{fig:ChangeOverTimeDaily:GooglePlayServices}
    \end{subfigure}
    \caption{User behavior change and profile drift for different Android applications}
    \label{fig:ChangeOverTimeDaily}
\end{figure}

Figure~\ref{fig:ChangeOverTimeDaily} shows how user behavior changes over time on the left side, and profile drift of each user on the right side for different Android applications.
The $x$-axis represents the day of the month, the $y$-axis represents the drift score for the user for that specific day compared to the aggregation of all activity in the previous days.

QWA allows its users to investigate reasons of the behavior changes by summarizing the features that caused the highest drift score change. This allows us to quickly inspect the information accessed when an alarm is raised. The trend line for each user represents the observed behavior drift, namely, how fast the behaviors of the user change. Less area under the trend line means the user is less inclined to change their daily routines.

In our dataset, There are 4 users who used Facebook app, and except one, the users have stable profiles. One user, on the other hand, seems to have a distinguishable behavior change over time. However, when inspected, that specific user only uses the application more than 3 minutes twice, which explains the spikes seen in Figure~\ref{fig:ChangeOverTimeDaily:Facebook}.

Similarly, except one user, all the users of Google+ application have steady behaviors. Most of the queries issued by Google+ app retrieve information on the user's account, which clearly shows how Android OS utilizes the Google+ application. Figure~\ref{fig:ChangeOverTimeDaily:GooglePlus}, on the other hand, reveals that this application is mostly affected by the phone usage characteristics.

%Hangouts, surprisingly, has the most adaptive user profiles. In Figure~\ref{fig:ChangeOverTimeDaily:Hangouts}, it can be seen that most user profiles have varying behaviors, which is understandable since all the other applications are used on a daily basis, but Hangouts application is a messaging platform which most people do not use everyday.

Google Play Services is an Android system-support app. The drift over time, as shown in Figure~\ref{fig:ChangeOverTimeDaily:GooglePlayServices}, is low for most of the users. This application controls the install, update, and delete application operations on behalf of the operating system. The inspection we performed shows that the user who has a distinguishing behavior drift is used to install, and delete various applications.

%Media Storage is another Android system-support application which mediates how the media files are stored on the phone. Again, except two users, the drift over time for most of the users is very small. However, there are two users who have very distinctive behaviors. When inspected, we saw that these two users use their phone as a music player.

One misconception from Figure~\ref{fig:ChangeOverTimeDaily} can be that similar trends in these graphs mean these users have analogous behavioral characteristics. Similar trends in these graphs only mean that the expectation of behavior change pace is comparable for these users. 

In the following section, we will describe the red-teaming approach we used to inject real workloads. Since these workloads were taken from the other users, the variety between users are directly correlated to the success of the experiments.

\subsection{Red-teaming approach}
\label{sec:redteaming}

%In this section, we describe the red-teaming approach to validate our framework's effectiveness on detecting attacks.
We consider two different attack frameworks: (1) Simulated query attack injection, in which we prepare specific attack scenarios for each app, and (2) Real workload injection, in which we input other users' real workloads into the user's own workload.
%\todo{We compare our system's true positive rate (TPR) against the false positive rate (FPR) to outlier detection model using various thresholds that do not consider temporal drift.}

From now on, we will call the actual workload owner \textit{the victim}, and the owner of the injected workload \textit{the adversary}.

\tinysection{Simulated workload injection} We inject specifically designed workloads to perform a malicious activity into the user workload.
We assume that all the actual query activity in the dataset is benign. 
Note that we inject the simulated workloads into the log, not into the actual smartphone databases.
%Android OS does not let other applications to access another application's database, and it is possible that they have other defense mechanisms in place against attacks.
%The attack scenarios given here does not imply we actually performed these attack's on our users' phones.
%In fact, if handwritten queries, which are not specifically designed to imitate the application generated queries, are injected into the workloads, they would be easier to distinguish from the preprogrammed queries.

This approach addresses free-styler and the first case of translator attack models described in Section~\ref{sec:threatmodel}:
\begin{itemize}
\item When a new workload consisting of queries that are not generated by the app's own query generator in addition to the benign workload is injected and run, it would equate to a free-styler attack since normally the app wouldn't produce these queries therefore not letting the victim the privilege to access the information.
\item In translator attack model, the malicious code modifies the query generation mechanism code of the app. The queries can still be generated by the app, but some of them will not reflect the same characteristics as the legitimate queries.
\end{itemize}
%This approach also indirectly addresses the external attacks that use the victim's credentials to access sensitive data or elevate the victim's privileges to access more information. This would reflect characteristics of masquerading and privilege elevation attacks.

The simulated queries we injected in the victim's workload are prepared according to the scenarios given below:
\begin{description}
\item[Facebook] We delete entries in the feed table and corresponding cache items. To replace them, we insert other feed items that we want the victim to believe.
\item[Google+] We access the account information and the photos stored on the account. We alter the account information in order to redirect the password renewal emails to us.
%\item[Hangouts] We access the locally stored messages and search for specific keywords to find sensitive information about the account owner. We insert the search results to the real time messages and send it to ourselves. We then remove the last message in order to cover our infiltration.
\item[Google Play Services] We modify the log records in order to confuse the operating system to skip updates for some applications. This would allow an adversary to take advantage of any patched vulnerabilities.
%\item[Media Storage] We remove the metadata information of media files and replace them with other irrelevant information aiming to cause the operating system to crash.
\end{description}

The results for this approach are given in Table~\ref{tab:synteticworkloadinjection}.

\begin{table}[]
\centering
\caption{Detection rates for profile drift using simulated workload injection }
\label{tab:synteticworkloadinjection}
\begin{tabular}{|c|c|c|c|c|}
\hline
                                         %                      & Users & \begin{tabular}[c]{@{}c@{}}\# of Attacks\\ Performed\end{tabular} & \begin{tabular}[c]{@{}c@{}}\# of Attacks\\ Detected\end{tabular} & \begin{tabular}[c]{@{}c@{}}Detection\\ Rate\end{tabular} \\ \hline
                                                                                    & \begin{tabular}[c]{@{}c@{}}\# of Attacks\\ Performed\end{tabular} & \begin{tabular}[c]{@{}c@{}}\# of Attacks\\ Detected\end{tabular} & \begin{tabular}[c]{@{}c@{}}Detection\\ Rate\end{tabular} \\ \hline
                                                                                     
Facebook                                 
%& 4    
& 105                                                               & 98                                                               & 93.3\%                                                   \\ \hline
Google+                                                       
%& 10
& 225                                                               & 214                                                              & 95.1\%                                                   \\ \hline
Google Play
%& 10 
& 282                                                               & 267                                                              & 94.7\%                                                   \\ \hline
\end{tabular}
\end{table}

\tinysection{Real workload injection} 
We inject different workloads created by other users into the normal workload of the user.
We assume that all the query activity in the dataset is benign. However, we simulate an attack by injecting one user's normal activity into the workload created by one of the other users.
Hence, we only use queries that were created by the app itself.

%We do not create any simulated workloads that could be viewed as biased.
%In fact, if handwritten queries, which are not specifically designed to imitate the application generated queries, are injected into the workloads, they would be easier to distinguish from the preprogrammed queries.

This approach addresses the copycat and the second case of the translator attack models described in Section~\ref{sec:threatmodel}:
\begin{itemize}
\item By injecting the adversary's workload, we simulate copycat attack. The adversary's workload would reflect its the characteristics of queries generated by legitimate actions while using the victim's credentials.
\item The second case of the translator attack similarly uses legitimate queries generated by the app, however, these queries are being generated by the modified code.
\end{itemize}
%As we discussed in Section~\ref{sec:threatmodel}, this approach also addresses the external attacks that use the victim's credentials to access sensitive data, namely, the attack equates to masquerading attacks.

The results for this approach are given in Table~\ref{tab:realworkloadinjection}.

\begin{table}[]
\centering
\caption{Detection rates for profile drift using real workload injection}
\label{tab:realworkloadinjection}
\begin{tabular}{|c|c|c|c|c|}
\hline
                                         %                      & Users & \begin{tabular}[c]{@{}c@{}}\# of Attacks\\ Performed\end{tabular} & \begin{tabular}[c]{@{}c@{}}\# of Attacks\\ Detected\end{tabular} & \begin{tabular}[c]{@{}c@{}}Detection\\ Rate\end{tabular} \\ \hline

& \begin{tabular}[c]{@{}c@{}}\# of Attacks\\ Performed\end{tabular} & \begin{tabular}[c]{@{}c@{}}\# of Attacks\\ Detected\end{tabular} & \begin{tabular}[c]{@{}c@{}}Detection\\ Rate\end{tabular} \\ \hline

Facebook                                 %                      & 4     
& 315                                                               & 283                                                              & 89.8\%                                                   \\ \hline
Google+                                   
%& 10
& 2025                                                              & 1818                                                             & 89.7\%                                                   \\ \hline
Google Play 
%& 10   
& 2583                                                              & 2092                                                             & 81.0\%                                                   \\ \hline
\end{tabular}
\end{table}

In our experiments, we created the normal user behavior model for each user. To compute detection accuracy rate, we partitioned the workloads of the same application created by all the other users day by day, and we tested each of these partitions of data to see if our system raises an alarm on the specific normal user behavior model that is being tested. As mentioned in Section~\ref{sec:redteaming}, although these injected workloads are actually naive workloads of other users, in our concept, they belong to a different user but still generated legitimately; hence, they represent an extremely skillful attacker. As shown in Table~\ref{tab:realworkloadinjection}, our methodology successfully determines these injections do not belong to the user model at least 81\% and at most 89.8\% of the times depending on the application.
%On the other hand, to compute the FPR, we rely on the actual behavior in the models that gets inaccurately classified as an attack which is actually just a naive outlier for the user model. Thus, the second and fourth columns of Tables~\ref{tab:synteticworkloadinjection} and~\ref{tab:realworkloadinjection} are computed by using data points that actually belongs to a specific user model. As an example, for Hangouts application, the behavior of a normal user tends to vary a lot over time, hence while static thresholding optimized for detecting injections determines 52\% of the user activity as an attack, our adaptive model only mistakes these behaviors 9\% of the time while still being able to detect injections as accurate as the other model. The FPR values reveal that we have a dramatic improvement in many cases over the existing strategies even if they seem to be high.

\section{Related Work}
\label{sec:relatedwork}
% !TEX root = ../paper.tex
There are two approaches to deal with data leakage from databases: (1) misuse detection, and (2) anomaly detection.

Misuse detection aims to collect a dataset of events that leads to intrusions.
These systems observe user behavior, and when a user takes certain actions, the system either raises an alarm, or blocks the user from taking any other actions.
These particular actions can be designed for specific scenarios to prevent well-known attacks, or they can be learned from other sources such as successfully caught incidents~\cite{Kul2015Jowua}.

Anomaly detection approach, on the other hand, depends on detecting anomalies at a user's behavior. The systems implemented with these approach can focus on a specific type of resource, or combination of resources~\cite{salem2008survey} such as file access patterns~\cite{Arief2015Jowua}, shell commands~\cite{nguyen2003systemcall}, and SQL queries~\cite{Kamra2007SyntaxBased, Mathew2010Raid, vavilis2015anomaly}.

%We focus on the anomaly detection approach on database systems to narrow down the enormous scope of the problem
There has been extensive level of research in detecting data leakage from databases, but there are still challenges in this field~\cite{Santos2014challenges}.
Chung \textit{et al.}~\cite{chung2000demids} proposed the use of access patterns to databases to detect typical behavior of users. Kamra \textit{et al.}~\cite{Kamra2007SyntaxBased} developed a SQL query feature extraction method that generalizes complex queries into simpler, and easier to compare forms to use them in detecting insider attacks. Mathew \textit{et al.}~\cite{Mathew2010Raid} introduced a data--centric approach that requires access to the data that a query returns, which then would be used to compute the overlap between returned result sets.
%Gafny \textit{et al.}~\cite{gafny2010context} proposed a similar framework to Mathew \textit{et al.}, adding the application context information on top of the detection mechanism, such as  location and user type.
%Costante \textit{et al.}~\cite{costante2013database} presented a scheme to find the root cause of data leakage in databases, and identifying the severity of the leak.
Wang \textit{et al.}~\cite{wang2010hengha} focused on harvesting attacks considering query correlation and result coverage.
Maggi \textit{et al.}~\cite{maggi2009protecting} is one of the leading works that introduced concept drift in web applications. Their model is designed to track the changes on websites in order to find out if there is a need to retrain the security application.

We use a temporal user behavior drift model in user profiles in this paper. To the best of our knowledge, our work is the first work that performs such an extensive study on SQL query data produced by real world users on temporal behavior drift. Although using temporal concept drift for outlier detection has been studied in a limited number of works before as pointed out, we believe these studies either did not use or create real-world user provided activities, or they used generated data. 
%One of the early works that introduced the temporal aspect in database workloads proposed that learning timing of different types of queries by specific users could improve intrusion detection rate~\cite{lee2000timesignatures}. This addresses queries that are issued out of working hours, or irregular updates to the database which usually take place regularly.
Our approach takes adaptation to individual behaviors into account which would allow flexibility to adapt to new tasks.

%Recently, detecting privacy leakage from smartphones has gained interest of both academia and industry. Many techniques have been proposed to identify leakage risk from apps in from different resources~\cite{hassanshahi2017android, jain2017sniffdroid, bhandari2017sneakleak}. However, these techniques mostly identify the risk, they do not directly address attacks. Our work bridges this gap by utilizing user habits in using mobile apps.

\section{Discussion}
\label{sec:discussion}
% !TEX root = ../paper.tex

%The model we described in this paper constitutes the first steps of building a temporal behavior drift prediction model. Concretely, we plan to extend our work in several directions: First, we will analyze other regression models, and test their effect on the performance of the system. Second, we incorporate a prediction model both for security applications, and performance optimization of database systems. Lastly, we will develop a production ready package to collect data in real world organizations, and test the system on site to eliminate privacy concerns.

There are a number of vulnerabilities that have been identified on Android OS which can lead to sensitive data leakage from the app database. In this paper, we discuss two of them, and argue that there may be other vulnerabilities with similar consequences that have not been discovered yet. The solution proposed in this paper is applicable to detect attacks that exploit these vulnerabilities. However, this method requires prior knowledge about how the user utilizes an app. Therefore, zero-day attacks cannot be detected with the proposed method. To address this problem, developers can insert probability distributions for different classes of users in the alpha tests of their app, so that the security layer that utilizes our method can collect enough data about the user, and still be able to detect zero day attacks.

In our experiments, we take the profile drift computation interval as one day, and we compare this distribution with the accumulated user profile over time. Although this approach achieves high detection rates in our setting, applying a sliding window on the streaming data to create the user profile can yield better results for different apps. The ideal length of the interval and sliding window depends on different settings.

%Another approach that can support the proposed method is clustering queries that represent the benign behavior of each user, and detecting individual queries that cannot be placed in any of the query clusters. Although this method achieves an average of 92\% on simulated workload injections for all the apps tested in this paper, it fails to detect more than 50\% of real workload injections. This is due to the fact that generated queries are constructed similarly by the same app, and injected queries are structurally similar to those of the user's.

Lastly, in this paper, we only propose a method to detect data leakage from databases on Android apps. We do not propose a strategy on how the app reacts when an attack is detected. We believe that this decision should be made by the app developer, and it is out of our scope. 

\section{Conclusions}
\label{sec:conclusions}
The focus of this paper was to highlight a class of vulnerabilities that can lead to data leakage from the database system, and to present a framework for creating user behavior profiles considering the temporal behavior drift to be used in detecting attacks exploiting them. We first provided a user behavior model for data leakage prevention. We argued that without considering the constant change in people's behaviors and habits, it is impossible for the defense systems to adapt to the new changes. This would result in the need for retraining of the user models for the system. In our experiments, we used real world query workloads and applied two different red teaming approaches: (1) simulated attack workloads, and (2) injection of benign real world workloads of other users.
%We evaluated our system with and without temporal behavior drift model in detecting these injections. For both approaches, the TPRs consistently remained comparable. However, FPR comparison revealed that temporal behavior drift model remarkably reduced the FPR, hence validated our assumptions.

The model we described in this paper constitutes the first steps of building a temporal behavior drift prediction model. Concretely, we plan to extend our work in several directions: First, we will analyze other regression models, and test their effect on the performance of the system. Second, we will incorporate a prediction model both for security applications, and performance optimization of database systems. Lastly, we will develop a production ready package to be integrated to apps and Android OS. %collect data in real world organizations, and test the system on site to eliminate privacy concerns.


\section*{Acknowledgment}
This material is based in part upon work supported by the National Science Foundation under award number CNS - 1409551. Usual disclaimers apply.



%\bibliographystyle{IEEEtran}

\balance

\bibliographystyle{IEEEtran}
\bibliography{paperGokhan}

\end{document}